\documentclass[runningheads]{llncs}

\usepackage{amssymb}
\usepackage{graphicx}

\usepackage{tipa}
\usepackage{t1enc}
\usepackage[latin1]{inputenc}
\usepackage{amsmath}
\usepackage{mathrsfs}
\usepackage{stmaryrd}
\usepackage{graphicx}
\usepackage{boxedminipage}
\usepackage{listings}
\usepackage{algpseudocode}
\usepackage[inline,nomargin,draft]{fixme}
\usepackage{xspace}

\newcounter{mscount}

\newdimen\proofrulebreadth \proofrulebreadth=.05em
\newdimen\proofdotseparation \proofdotseparation=1.25ex
\newdimen\proofrulebaseline \proofrulebaseline=2ex
\newcount\proofdotnumber \proofdotnumber=3
\let\then\relax
\def\hfi{\hskip0pt plus.0001fil}
\mathchardef\squigto="3A3B
\newif\ifinsideprooftree\insideprooftreefalse
\newif\ifonleftofproofrule\onleftofproofrulefalse
\newif\ifproofdots\proofdotsfalse
\newif\ifdoubleproof\doubleprooffalse
\let\wereinproofbit\relax
\newdimen\shortenproofleft
\newdimen\shortenproofright
\newdimen\proofbelowshift
\newbox\proofabove
\newbox\proofbelow
\newbox\proofrulename
\def\shiftproofbelow{\let\next\relax\afterassignment\setshiftproofbelow\dimen0 }
\def\shiftproofbelowneg{\def\next{\multiply\dimen0 by-1 }%
\afterassignment\setshiftproofbelow\dimen0 }
\def\setshiftproofbelow{\next\proofbelowshift=\dimen0 }
\def\setproofrulebreadth{\proofrulebreadth}

\def\prooftree{
\ifnum	\lastpenalty=1
\then	\unpenalty
\else	\onleftofproofrulefalse
\fi
\ifonleftofproofrule
\else	\ifinsideprooftree
	\then	\hskip.5em plus1fil
	\fi
\fi
\bgroup
\setbox\proofbelow=\hbox{}\setbox\proofrulename=\hbox{}%
\let\justifies\proofover\let\leadsto\proofoverdots\let\Justifies\proofoverdbl
\let\using\proofusing\let\[\prooftree
\ifinsideprooftree\let\]\endprooftree\fi
\proofdotsfalse\doubleprooffalse
\let\thickness\setproofrulebreadth
\let\shiftright\shiftproofbelow \let\shift\shiftproofbelow
\let\shiftleft\shiftproofbelowneg
\let\ifwasinsideprooftree\ifinsideprooftree
\insideprooftreetrue
\setbox\proofabove=\hbox\bgroup$\displaystyle 
\let\wereinproofbit\prooftree
\shortenproofleft=0pt \shortenproofright=0pt \proofbelowshift=0pt
\onleftofproofruletrue\penalty1
}

\def\eproofbit{
\ifx	\wereinproofbit\prooftree
\then	\ifcase	\lastpenalty
	\then	\shortenproofright=0pt	
	\or	\unpenalty\hfil		
	\or	\unpenalty\unskip	
	\else	\shortenproofright=0pt	
	\fi
\fi
\global\dimen0=\shortenproofleft
\global\dimen1=\shortenproofright
\global\dimen2=\proofrulebreadth
\global\dimen3=\proofbelowshift
\global\dimen4=\proofdotseparation
\setcounter{mscount}{\proofdotnumber}
$\egroup  
\shortenproofleft=\dimen0
\shortenproofright=\dimen1
\proofrulebreadth=\dimen2
\proofbelowshift=\dimen3
\proofdotseparation=\dimen4
\proofdotnumber=\value{mscount}
}

\def\proofover{
\eproofbit 
\setbox\proofbelow=\hbox\bgroup 
\let\wereinproofbit\proofover
$\displaystyle
}%
\def\proofoverdbl{
\eproofbit 
\doubleprooftrue
\setbox\proofbelow=\hbox\bgroup 
\let\wereinproofbit\proofoverdbl
$\displaystyle
}%
\def\proofoverdots{
\eproofbit 
\proofdotstrue
\setbox\proofbelow=\hbox\bgroup 
\let\wereinproofbit\proofoverdots
$\displaystyle
}%
\def\proofusing{
\eproofbit 
\setbox\proofrulename=\hbox\bgroup 
\let\wereinproofbit\proofusing
\kern0.3em$
}

\def\endprooftree{
\eproofbit 
  \dimen5 =0pt
\dimen0=\wd\proofabove \advance\dimen0-\shortenproofleft
\advance\dimen0-\shortenproofright
\dimen1=.5\dimen0 \advance\dimen1-.5\wd\proofbelow
\dimen4=\dimen1
\advance\dimen1\proofbelowshift \advance\dimen4-\proofbelowshift
\ifdim	\dimen1<0pt
\then	\advance\shortenproofleft\dimen1
	\advance\dimen0-\dimen1
	\dimen1=0pt
	\ifdim  \shortenproofleft<0pt
        \then   \setbox\proofabove=\hbox{%
			\kern-\shortenproofleft\unhbox\proofabove}%
                \shortenproofleft=0pt
        \fi
\fi
\ifdim	\dimen4<0pt
\then	\advance\shortenproofright\dimen4
	\advance\dimen0-\dimen4
	\dimen4=0pt
\fi
\ifdim	\shortenproofright<\wd\proofrulename
\then	\shortenproofright=\wd\proofrulename
\fi
\dimen2=\shortenproofleft \advance\dimen2 by\dimen1
\dimen3=\shortenproofright\advance\dimen3 by\dimen4
\ifproofdots
\then
	\dimen6=\shortenproofleft \advance\dimen6 .5\dimen0
	\setbox1=\vbox to\proofdotseparation{\vss\hbox{$\cdot$}\vss}%
	\setbox0=\hbox{%
		\advance\dimen6-.5\wd1
		\kern\dimen6
		$\vcenter to\proofdotnumber\proofdotseparation
			{\leaders\box1\vfill}$%
		\unhbox\proofrulename}%
\else	\dimen6=\fontdimen22\the\textfont2 
	\dimen7=\dimen6
	\advance\dimen6by.5\proofrulebreadth
	\advance\dimen7by-.5\proofrulebreadth
	\setbox0=\hbox{%
		\kern\shortenproofleft
		\ifdoubleproof
		\then	\hbox to\dimen0{%
			$\mathsurround0pt\mathord=\mkern-6mu%
			\cleaders\hbox{$\mkern-2mu=\mkern-2mu$}\hfill
			\mkern-6mu\mathord=$}%
		\else	\vrule height\dimen6 depth-\dimen7 width\dimen0
		\fi
		\unhbox\proofrulename}%
	\ht0=\dimen6 \dp0=-\dimen7
\fi
\let\doll\relax
\ifwasinsideprooftree
\then	\let\VBOX\vbox
\else	\ifmmode\else$\let\doll=$\fi
	\let\VBOX\vcenter
\fi
\VBOX	{\baselineskip\proofrulebaseline \lineskip.2ex
	\expandafter\lineskiplimit\ifproofdots0ex\else-0.6ex\fi
	\hbox	spread\dimen5	{\hfi\unhbox\proofabove\hfi}%
	\hbox{\box0}%
	\hbox	{\kern\dimen2 \box\proofbelow}}\doll%
\global\dimen2=\dimen2
\global\dimen3=\dimen3
\egroup 
\ifonleftofproofrule
\then	\shortenproofleft=\dimen2
\fi
\shortenproofright=\dimen3
\onleftofproofrulefalse
\ifinsideprooftree
\then	\hskip.5em plus 1fil \penalty2
\fi
}

\newcommand{\reglaName}[1]{\ensuremath{\scriptscriptstyle{\text{#1}}}}
\newcommand{\regla}[3]{\dfrac{#1}{#2}\reglaName{(#3)}}

\newcommand{\pest}{\textsc{Pest}\xspace}
\newcommand{\budapest}{\textsc{BudaPest}\xspace}
\newcommand{\pestFuente}[1]{\textbf{#1}}

\newcommand{\atpre}{\pestFuente{@pre}}
\newcommand{\expVarAtPre}[1]{#1\atpre}

\newcommand{\expArraySize}[1]{|#1|}
\newcommand{\expArrayAccess}[2]{#1[#2]}
\newcommand{\expNumeral}[1]{#1}

\newcommand{\expTrue}{\pestFuente{true}}

\newcommand{\expNot}[1]{\lnot #1}
\newcommand{\expQuantif}[3]{#1\ #2\ (#3)}
\newcommand{\expQuantifAcotada}[7]{#1#4\ \slash\ #2 #3 #4 #5 #6:\ #7}
\newcommand{\expExists}{\exists}
\newcommand{\expForall}{\forall}

\newcommand{\sentAssign}[2]{#1 \ensuremath{\leftarrow} #2}
\newcommand{\sentArrayPositionAssign}[3]{#1[#2] \ensuremath{\leftarrow} #3}
\newcommand{\sentLocalDef}[3]{\pestFuente{local}\ #1\ensuremath{\leftarrow} #3}
\newcommand{\sentIfThenElse}[3]{\pestFuente{if}\ #1\ \pestFuente{then}\ #2\ \pestFuente{else}\ #3\ \pestFuente{fi}}
\newcommand{\sentWhile}[4]{\pestFuente{while}\ #1\ \pestFuente{:?!}\ #2\ \pestFuente{:\#}\ #3\ \pestFuente{do}\ #4\ \pestFuente{od}}

\newcommand{\sentSkip}{\pestFuente{skip}}
\newcommand{\sentCall}[2]{\pestFuente{call}\ #1(#2)}

\newcommand{\sentMap}[3]{\pestFuente{map}\ #1\ \pestFuente{in}\ #2[\pestFuente{..}\ #3\ \pestFuente{..}]}

\newcommand{\sentFor}[4]{\pestFuente{for}\ #1\ \pestFuente{from}\ #2\ \pestFuente{to}\ #3\ \pestFuente{do}\ #4\ \pestFuente{od}}

\newcommand{\procPre}[1]{\operatorname{pre}(#1)}
\newcommand{\procPost}[1]{\operatorname{post}(#1)}
\DeclareMathOperator{\procBody}{body}

\newcommand{\progEmpty}{\emptyset}
\newcommand{\program}{\pi}

\algdef{LS}{Procedure}{0}[2]{#1(#2)}
\algdef{LC}{Procedure}{Pre}{}[1]{\pestFuente{:?}\ #1}
\algdef{LC}{Pre}{Post}{}[1]{\pestFuente{:!}\ #1}
\algdef{CE}{Post}{Begin}{End}{\{}{\}}

\algdef{LS}{Skip}{0}{\sentSkip}
\algdef{LS}{Assign}{0}[2]{\sentAssign{#1}{#2}}
\algdef{LS}{LocalDef}{0}[3]{\sentLocalDef{#1}{#2}{#3}}
\algdef{LS}{ArrayPositionAssign}{0}[3]{\sentArrayPositionAssign{#1}{#2}{#3}}
\algdef{LS}{LocalDefArray}{0}[2]{\sentLocalDefArray{#1}{#2}}

\algdef{SE}{If}{Fi}[1]{\pestFuente{if}\ #1\ \pestFuente{then}}{\pestFuente{fi}}
\algdef{CE}{If}{Else}{Fi}{\pestFuente{else}}{\pestFuente{fi}}

\algdef{LS}{While}{2}[1]{\pestFuente{while}\ #1}
\algdef{LS}{Inv}{}[1]{\pestFuente{:?!}\ #1}
\algdef{LC}{Inv}{Var}{0}[1]{\pestFuente{:\#}\ #1}
\algdef{CE}{While}{Do}{Od}{\pestFuente{do}}{\pestFuente{od}}

\algdef{SE}{Map}{In}{\pestFuente{map}}[2]{\pestFuente{in}\ #1[\pestFuente{..}\ #2\ \pestFuente{..}\hspace*{-.25em}]}
\algdef{SE}{Find}{In}[2]{#1 \ensuremath{\leftarrow} \pestFuente{find}\ #2\ \pestFuente{as}}[2]{\pestFuente{in}\ #1[\pestFuente{..}\ #2\ \pestFuente{..}\hspace*{-.25em}]}
\algdef{SE}{For}{Od}[3]{\pestFuente{for}\ #1\ \pestFuente{from}\ #2\ \pestFuente{to}\ #3\ \pestFuente{do}}{\pestFuente{od}}

\algdef{LS}{Call}{0}[2]{\sentCall{#1}{#2}}%

\DeclareMathOperator{\modVars}{modVars}
\DeclareMathOperator{\localVars}{locals}

\newcommand{\clasifInt}{\textsc{Int}}

\newcommand{\clasifBool}{\textsc{Bool}}

\DeclareMathOperator{\safe}{safe}
\newcommand{\safeExp}[1]{\safe}
\DeclareMathOperator{\safeBoolExp}{\safeExp{\clasifBool}}

\DeclareMathOperator{\safeIntExp}{\safeExp{\clasifInt}}

\DeclareMathOperator{\memFunc}{\sigma}

\DeclareMathOperator{\pmemFunc}{\rho}

\newcommand{\memExt}[3]{\funcExtend{#1}{#2}{#3}}

\newcommand{\memPoda}[2]{#1 \hspace*{-.1em} \ominus #2}

\newcommand{\semExp}[3]{\llbracket #2 \rrbracket_{#1}} 
\newcommand{\semintExp}[2]{\semExp{#1}{#2}{\clasifInt}}

\newcommand{\semboolExp}[2]{\semExp{#1}{#2}{\clasifBool}}
\newcommand{\semSent}[3]{#1\ \triangleright\ #2\ \triangleright\ #3}

\newcommand{\as}[3]{\set{#1}\ #2\ \set{#3}}

\newcommand{\clausuraExistencial}[2]{Cl_{\exists #2}(#1)}

\newcommand{\calcPost}[2]{\operatorname{post}(#1,#2)}
\newcommand{\calcPre}[2]{\operatorname{pre}(#1,#2)}

\newcommand{\trSent}[2]{\operatorname{tr^{S}}(#1,\ #2)}
\newcommand{\trSentBig}[2]{\operatorname{tr^{S}}\Bigl(#1,\ #2\Bigr)}

\newcommand{\funcExtend}[3]{\ensuremath{#1\lbrace #2 \mapsto #3\rbrace}}

\DeclareMathOperator{\ltrue}{true}
\DeclareMathOperator{\lfalse}{false}

\newcommand{\fuerza}[2]{#1 \models #2}

\newcommand{\set}[1]{\ensuremath{\lbrace #1 \rbrace}}

\newcommand{\sustPart}[2]{\ensuremath{#1 \mapsto #2}}
\newcommand{\sustFrame}[2]{\ensuremath{#1\lfloor#2\rfloor}}
\newcommand{\sust}[3]{\sustFrame{#1}{\sustPart{#2}{#3}}}

\newcommand{\eqdef}{\ensuremath{\overset{\text{def}}{=}}}

\newcommand{\findcite}[1]{\cite{estoy-cagado}}

\newtheorem{teoremaAux}{Theorem}

\newtheorem{thm}{Theorem}
\newtheorem{defn}{Definition}

\newcommand{\algoritmo}[3]{
\label{alg:#1}
\begin{center}
\begin{minipage}{#2}
\begin{algorithmic}[0]
#3
\end{algorithmic}
\end{minipage}
\end{center}
}

\usepackage{url}
\urldef{\mailsa}\path|{alfred.hofmann,ursula.barth,ingrid.beyer,natalie.brecht,|
\urldef{\mailsb}\path|christine.guenther,frank.holzwarth,piamaria.karbach,|
\urldef{\mailsc}\path|anna.kramer,erika.siebert-cole,lncs}@springer.com|
\newcommand{\keywords}[1]{\par\addvspace\baselineskip
\noindent\keywordname\enspace\ignorespaces#1}

\begin{document}

\mainmatter  

\title{Reducing the Number of Annotations \\
in a Verification-oriented Imperative Language}

\titlerunning{Reducing Annotations in a Verification-oriented Language}

%
%
\author{Guido de Caso\and Diego Garbervetsky\and Daniel Gorín}
\authorrunning{de Caso, Garbervetsky, Gorín}

\institute{Departamento de Computaci{\'o}n, FCEyN, Universidad de Buenos Aires \\
\{gdecaso,diegog,dgorin\}@dc.uba.ar}

\toctitle{Reducing the Number of Annotations in a Verification-oriented Imperative Language}
\tocauthor{de Caso, Garbervetsky, Gorín}
\maketitle

\begin{abstract}
Automated software verification is a very active field of research which has made enormous progress both in theoretical and practical aspects. Recently, an important amount of research effort has been put into applying these techniques on top of mainstream programming languages. These languages typically provide powerful features such as reflection, aliasing and polymorphism which are handy for practitioners but, in contrast, make verification a real challenge.
In this work we present \pest, a simple experimental, while-style, multiprocedural, imperative programming language which was conceived with verifiability as one of its main goals. 
This language forces developers to concurrently think about both the statements needed to implement an algorithm and the assertions required to prove its correctness. 
In order to aid programmers, we propose several techniques to reduce the number and complexity of annotations required to successfully verify their programs. In particular, we show that \emph{high-level iteration constructs} 
may alleviate the need for providing complex loop annotations.
\end{abstract}

\section{Introduction}

Formal automated software verification regained in recent years the attention of the community. There are at least two reasons behind this resurgent success: on the one hand, there were crucial developments in automated theorem proving in the last fifteen years, with \textsc{SAT}-solvers finally reaching industrial strength; on the other, the focus was shifted to \emph{partial specifications} which somehow overcomes many of the objections raised in~\cite{de-millo}. Verification of partial specifications must be regarded as an error-detection procedure and, as such, akin to traditional forms of testing.

We will center on the particular form of verification where the source code is annotated with special assertions. These normally take the form of method pre and postconditions, loop and class invariants, etc.  Special tools then read these annotated sources, generate verification conditions from them and feed these into automated provers~\cite{vcs}.  \textsc{Spec\#}~\cite{spec-sharp-overview}
and \textsc{ESC/Java}~\cite{esc-java} are two of the best-known examples. The former is based on a dialect of \textsc{C\#} while the latter takes \textsc{Java} code with \textsc{JML}~\cite{jml} annotations.

Apart from the abovementioned, there are ongoing research efforts in automated verification for almost every major programming language in use~\cite{havoc,esc-haskell,pychecker}. The rationale is to lower the adoption barrier by giving practitioners tools for verifying the code they are writing today. Now, while this is an undeniably sensible plan, we perceive that, in almost every case, the resulting ``programming-language-with-annotations'' regarded as a whole ends-up being not entirely satisfying. We consider next some of the reasons for this.

\paragraph{Lack of cohesion.} Annotations are usually introduced as a ``patch'' to the language. Most of the time, this is done in a way such that regular compilers and \textsc{IDE}-tools regard them as mere comments. Moreover,  most programming languages provide a way to perform \emph{optional run-time assertion checks}. These are usually used to validate pre and postconditions or invariants and are, thus, the run-time counterparts of the verification annotations. But despite their dual nature, both mechanisms have normally no syntactical relation whatsoever. 

\paragraph{Redundancy.} In statically typed languages, the type of the input and output variables of a function are clearly part of its contract. But this means one ends-up with two completely unrelated ways of specifying contracts: one enforced by compilers (types) and the other by static checkers (the additional annotations). 

\paragraph{Missed optimization opportunities.} Optimizing compilers cannot leverage on program annotations in the same way they currently do on type information.

\paragraph{Inadequate semantics.} We can most certainly exclude \emph{``to ease automated verifiability''} from the list of goals that have driven the design of most modern-day programming languages. We cannot know for sure if today's mainstream languages would have been as popular without features such as complex inheritance mechanisms, uncontrolled method reentrancy or unrestricted aliasing. Nevertheless, the fact that the designers of \textsc{Spec\#} already had to diverge in slight ways from \textsc{C\#}'s semantics~\cite{spec-sharp-overview} is indicating, in our opinion, a new driver for the programming languages to come. 

\smallskip
In this paper we report on an ongoing experiment in language design. 
Our language forces programmers to introduce annotations specifying their intention when writing code.
We will show that by coupling tightly the annotations with the language semantics we can rely on a verifying compiler to infer many of the annotations. Moreover, we find that \emph{high level iteration constructs} can effectively be used as means of reducing the annotation burden of invariants.

\section{A walk-through of the  \pest language}\label{sec:motivation}

In this section we will briefly present the \pest\footnote{After the eastern side of the Danube river in Budapest, pronounced [\textipa{"pESt}].} programming language by way of examples. For a formal description the reader is referred to~\cite{pest}.

\pest is an multi-procedural, non-recursive, structured, while-style, imperative programming language whose syntax natively incorporates various kinds of annotations. Its main objective is to provide a test-bed for exploring new concepts and ideas.

\begin{figure}
\begin{lstlisting}[emph={max}]
  max(a,b,c)
  :? true
  :! (a >= b => c = a) && (a < b => c = b)
  :! a = a@pre && b = b@pre
  {
    if a >= b then
      c <- a
    else
      c <- b
    fi
  }
\end{lstlisting}
\caption{Simple \pest procedure definition.}
\label{fig:max}
\end{figure}

Figure~\ref{fig:max} shows the definition of a simple procedure in \pest.
Keywords~\lstinline{:?} and~\lstinline{:!} introduce pre and postconditions respectively.
In the postcondition \lstinline{a@pre} and \lstinline{b@pre} denote the value of \lstinline{a} and \lstinline{b} at the beginning of the procedure execution. Therefore, this clause states that the value of \lstinline{a} and \lstinline{b} does not change; this is necessary since in \pest all values are copied-in and then copied-out.

There are currently only three data types in \pest: booleans, integers and arrays of integers. Variables are monomorphic and their type is inferred from use. For example, in Figure~\ref{fig:max}, all variables are integers, since the \lstinline{>=} operator takes integers as arguments.  Apart from classical boolean operators,  boolean expressions may include \emph{bounded} first-order quantification.

Figure~\ref{fig:arrayMax} lists a  \pest procedure containing a while-loop iterating over an array and a procedure call. Loop invariants are introduced using the \lstinline{:?!} keyword and exactly one loop variant must be provided, using the \lstinline{:#} keyword. Only variables are allowed as arguments on procedure calls and they are syntactically enforced to be distinct. Observe that since, additionally, values are copied in variable assignments, we impose a strict control over aliasing.

\begin{figure}[t]
\begin{lstlisting}[emph={max,arrayMax}]
arrayMax(A, m)
:? |A| > 0
:! forall-k / 0 <= k < |A| : m >= A[k]
:! exists-k / 0 <= k < |A| : m = A[k]
:! A = A@pre
{
  m <- A[0]
  local i <- 1
  while i < |A|
      :?! 1 <= i <= |A|
      :?! forall-k / 0 <= k < i : m >= A[k]
      :?! exists-k / 0 <= k < i : m = A[k]
      :?! A = A@pre
      :# |A| - i
   do
     local t <- 0
     local e <- A[i]
     call max(e, m, t)
     m <- t
     i <- i + 1
   od
}
\end{lstlisting}
\caption{A \pest procedure containing a loop.}
\label{fig:arrayMax}
\end{figure}

Operationally, annotations in \pest are interpreted as assertions. 
Roughly speaking, preconditions are evaluated prior to procedure calls and postconditions on return; invariants are checked before evaluating the loop guard. 
Since most of the semantic rules include the evaluation of annotations, a program execution gets stuck when the interpretation of one of them fails.

To enforce this semantics, annotations can be checked at run-time, but this can be very expensive. Alternatively, a \pest compiler can simply remove an assertion if it can be statically verified. In a way, this is reminiscent of \emph{type erasure}. Of course, the programmer will normally want to be aware of which assertions failed to be statically verified.

A \pest compiler can also \emph{infer} pre and postconditions of a procedure. If only the precondition is given a postcondition can be obtained by way of a symbolic computation; on the other hand, starting from a postcondition, a precondition can be obtained using a variation of the notion of \emph{weakest precondition}. In simple cases, like the procedure in Figure~\ref{fig:max}, one can remove the annotation altogether and rely solely on inference. Details are given in \S\ref{sec:burden}.  

Eliminating the need for explicitly given loop invariants is, of course, highly desirable. A lot of research has been done in loop invariant inference (for instance, \cite{daikon,leinoL05}). 
In \S\ref{sec:constructs} we will show that higher-level iteration constructs can be alternatively regarded as encapsulating ``directives'' on how to build a proper invariant from the loop body.

Using the ideas sketched above, \lstinline[emph={arrayMax}]{arrayMax} can be rewritten in \pest as shown in Figure~\ref{fig:easyArrayMax} which requires no annotations for the loop and still enforces its contract.
Notice that, in addition to the invariant, in this case we are also able to remove the precondition and part of the postcondition since they can be inferred by the \pest compiler using the rules presented in \S\ref{sec:burden}.

\begin{figure}
\begin{lstlisting}[emph={max,easyArrayMax}]
easyArrayMax(A,m)
:! forall-k / 0 <= k < |A| : m >= A[k]
:! exists-k / 0 <= k < |A| : m = A[k]
{
    m <- A[0]
    for i from 1 to |A| do
        local t <- 0
        local e <- A[i]
        call max(e, m, t)
        m <- t
    od
}
\end{lstlisting}
\caption{Using a \lstinline{for} construct to remove annotations.}
\label{fig:easyArrayMax}
\end{figure}

\section{Formal semantics of \pest}\label{sec:semantics}

In this section we  introduce \pest  semantics.
In \S\ref{subsec:oper} we briefly comment on \pest (big-step) operational semantics. As we already mentioned in the preceding section, annotations are interpreted as assertions and a computation gets stuck whenever one of these fails.
In \S\ref{subsec:hoare}, we present a static formulation based on Hoare-style clauses. This characterizes those programs that cannot get stuck according to the operational semantics. In this sense, it is reminiscent of a type system;  but, observe that it relies on a semantic entailment relation over formulas ($\models$) that is undecidable in the general case.

In this setting, automatic verification of a \pest program can be performed by checking conformance to its static semantics using a computable approximation ($\vdash$) of the $\models$-relation. Later, in \S\ref{sec:burden} we will derive some calculi from the static semantics to cope with several inference tasks.

\subsection{Operational semantics}\label{subsec:oper}

\pest \emph{operational semantics} is based on the standard notion of while-style programming languages semantics. Annotations are handled by incorporating \emph{conditions} (i.e. assertions) into the semantic rules that  disallow the execution of a program that violates them.

The semantic rules are given in terms of state transformations. A state $\memFunc$ is a function that maps program variables to concrete values of the proper type. With 
$\memExt{\memFunc}{v}{n}$ we denote a state that coincides with $\memFunc$ except, possibly, in the value for $v$ which, in the former, is $n$; $\memPoda{\memFunc}{V}$ is the restriction of $\memFunc$ to a domain that does not contain any variable in $V$. We use $\semExp{\memFunc}{e}{}$ to denote the \emph{value} of an expression $e$ under $\memFunc$.

We write $\semSent{\memFunc}{s}{\memFunc'}$ to express that a \pest sentence $s$, when run from state $\memFunc$, finishes correctly and arrives in state $\memFunc'$.
Some rules are depicted in Figure~\ref{fig-semantics}; for the complete set of rules refer to~\cite{pest}.

\begin{figure}[ht!]
\begin{small}
$$
\begin{array}{c}
\regla{\semboolExp{\memFunc}{g} = \lfalse \quad \semboolExp{\memFunc}{inv} = \ltrue}{\semSent{\memFunc}{\sentWhile{g}{inv}{var}{s}}{\memFunc}}{O-WHILE-F}
\\
\\
\regla{\begin{array}{c}\semboolExp{\memFunc}{g} = \ltrue \quad \semboolExp{\memFunc}{inv} = \ltrue \\
\semintExp{\memFunc}{var} > 0 \quad \semSent{\memFunc}{s}{\memFunc'} \\
\semintExp{\memFunc'}{var} < \semintExp{\memFunc}{var} \\
\semSent{\memPoda{\memFunc'}{\localVars(s)}}{\sentWhile{g}{inv}{var}{s}}{\memFunc''}\end{array}}
{\semSent{\memFunc}{\sentWhile{g}{inv}{var}{s}}{\memFunc''}}{O-WHILE-T}
\\
\\
\regla{\begin{array}{c}\semboolExp{\pmemFunc}{\procPre{proc}} = \ltrue \\
\semSent{\pmemFunc}{\procBody(proc)}{\pmemFunc'}\\
\semboolExp{\pmemFunc'}{\procPost{proc}} = \ltrue\end{array}}{\semSent{\memFunc}{\sentCall{proc}{cp_1,\ldots,cp_k}}{\memFunc\{\mathit{cp}_1 \mapsto \pmemFunc'(p_1),\ldots\}}}{O-CALL}
\\
\text{where}\ \pmemFunc(p_i) \eqdef \memFunc(cp_i)\ \text{and}\ \pmemFunc(\expVarAtPre{p_i}) \eqdef \memFunc(cp_i)
\end{array}
$$
\end{small}
\caption{\pest operational semantics (fragment)}
\label{fig-semantics}
\end{figure}

Consider the rules for the \pestFuente{while} statement. Observe first that if $\semboolExp{\memFunc}{inv} \neq \ltrue$ then no rule applies and, thus, a computation will stick in that case. When the guard is false, the state is not affected (the language syntax guarantees that guards are free of side-effects). Alternatively, when the guard is true, the variant must be above zero in order to continue. Observe that $\memFunc'$ is the state after an execution of the loop body, including locally-defined variables; if the variant didn't decrease the computation will stuck.

For procedure calls, state $\pmemFunc$ binds formal and actual parameters; the precondition of the called procedure must hold for this state.  The state $\pmemFunc'$, if defined, is the result of executing the procedure body from $\pmemFunc$; the postcondition must hold in $\pmemFunc'$. At the end, actual parameters are updated with the final values assigned to the formal parameters (recall that all parameters are in/out in \pest).

\subsection{Hoare-style static semantics}\label{subsec:hoare}

We first need to introduce some preliminary definitions. For $\memFunc$ a state and $b$ a \pest boolean expression\footnote{Throughout this section, boolean expressions are augmented with unbounded existential quantification.}, we write $\fuerza{\memFunc}{b}$ if $\semExp{\memFunc}{b}{} = \ltrue$. We will write $\fuerza{b_1}{b_2}$ to indicate that, for all $\memFunc$, whenever  $\fuerza{\memFunc}{b_1}$ holds, then  $\fuerza{\memFunc}{b_2}$ must hold too (i.e. $b_1$ is \emph{stronger} than $b_2$).

We also require a notion of ``safeness'' for  the evaluation of expressions. Given an expression $e$, we want $\safeExp{\sigma}(e)$ to be a boolean expression such that $\fuerza{\memFunc}{\safeExp{\sigma}(e)}$ implies that $\semExp{\memFunc}{e}{}$ is defined.  For example, we expect $\safeIntExp(\mbox{\lstinline{a[i] / y}})$ to be the expression ``\lstinline{0 <= i && i < |a| && y /= 0}''. A formal definition of these conditions is straightforward. Finally, $\sust{e_1}{e_2}{e_3}$ denotes the expression that results from replacing every occurrence of $e_2$ by $e_3$ in $e_1$. Observe that $e_2$ and $e_3$ must be of the same type.

Instead of using states like in the operational case, for the static semantics we will use predicates (i.e. boolean expressions) that describe a (possibly infinite) set of states. For boolean expressions $p$ and $q$ and $s$ a \pest sentence, $\as{p}{s}{q}$ must be read ``after executing $s$ from a $\memFunc$ such that $\fuerza{\memFunc}{p}$, we obtain a $\memFunc'$ such that $\fuerza{\memFunc'}{q}$''.

Figure~\ref{fig-abstract-semantics} lists some of the static rules (for the complete list, refer to~\cite{pest}). Consider the rule for assignments: the first premise guarantees that the program won't get stuck when evaluating $e$; the second one states that $q$ is a consequence of what was known prior to the assignment ($\sust{p}{v}{v'}$) and its  effect ($v=\sust{e}{v}{v'}$). The existentially quantified variable $v'$ stands for the value of $v$ before the assignment (this requires unbounded quantification).

\begin{figure}
\begin{small}
$$
\begin{array}{c}
\regla{\begin{array}{c}\fuerza{p}{\safeExp{\sigma}(e)} \\ \fuerza{\expQuantif{\expExists}{v'}{\sust{p}{v}{v'}\ \land\ v = \sust{e}{v}{v'}}}{q}\end{array}}{\as{p}{\sentAssign{v}{e}}{q}}{S-ASSIGN}
\\
\\
\regla{\begin{array}{c}
\fuerza{p}{\safeBoolExp(g)} \\
\fuerza{p \land g}{p_1} \quad \as{p_1}{s_1}{q_1} \quad \fuerza{q_1}{q} \\
\fuerza{p \land \expNot{g}}{p_2} \quad \as{p_2}{s_2}{q_2} \quad\fuerza{q_2}{q}\end{array}}
{\as{p}{\sentIfThenElse{g}{s_1}{s_2}}{q}}{S-IF}
\\
\\
\regla{\begin{array}{c}
\fuerza{\expTrue}{\safeBoolExp(inv)} \quad \fuerza{inv}{\safeIntExp(var)} \\
\fuerza{inv}{\safeBoolExp(g)} \quad \fuerza{p}{inv} \quad \fuerza{inv \land g}{p'} \\
\fuerza{p'}{var > \expNumeral{0}} \quad \as{p'}{\sentAssign{var_{0}}{var}\;\; s}{q'} \\
\fuerza{q'}{inv} \quad \fuerza{q'}{var < var_{0}} \quad \fuerza{inv \land \expNot{g}}{q}\end{array}}
{\as{p}{\sentWhile{g}{inv}{var}{s}}{q}}{S-WHILE}
\end{array}
$$
\end{small}
\caption{\pest static semantics (fragment)}
\label{fig-abstract-semantics}
\end{figure}

The premises of the rule for \pestFuente{while} can be seen as both a proof of the Fundamental Invariance Theorem for Loops~\cite{discipline-dijkstra} and a proof of termination using the loop variant. The predicate $p'$ represents any state where the invariant and the condition of the while hold; the loop body is augmented with an initial assignment to a \emph{fresh} variable $\mathit{var}_0$ that is used to prove that the variant decreases.

There is a clear correlation between \pest's operational and static semantics. Using the latter, we can give a notion of \emph{safe} program. In what follows, if $\pi$ is a program and $p$ a procedure, then $\pi, p$ is the program obtained by appending $p$ to $\pi$.

\begin{defn}[Safe programs]
The set \textsc{Safe} of programs is inductively defined as follows:
$$
\regla{}{\progEmpty \in \text{\textsc{Safe}}}{SAFE-EMPTY}
$$
$$
\regla{\program \in \text{\textsc{Safe}} \quad \as{\procPre{p}}{\procBody(p)}{\procPost{p}}}{\program, p \in \text{\textsc{Safe}}}{SAFE-EXTEND}
$$
\end{defn}

\begin{thm}[Safe programs execute normally]\label{programas_seguros_sem}
Let $\program \in \mbox{\textsc{Safe}}$ and let $p$ be a procedure in $\program$. For each $\memFunc$ such that $\fuerza{\memFunc}{\procPre{p}}$, there exists a state $\memFunc'$ such that $\semSent{\memFunc}{\procBody(p)}{\memFunc'}$ and $\fuerza{\memFunc'}{\procPost{p}}$.
\end{thm}
\vspace{-3mm}
\begin{proof}
It is a longish yet straightforward induction on the length of a derivation of $\as{\procPre{p}}{\procBody(p)}{\procPost{p}}$. See~\cite{pest}.
\end{proof}

\section{Reducing the annotation burden}\label{sec:burden}
In the previous section we presented the \pest programming language, including a soundness result showing that programs conforming to \pest static semantics do not go wrong. Still, providing \pest annotations is a heavy and complex task. In this section we discuss two techniques that assist the programmer by inferring and completing annotations.

\subsection{Inference of procedure contracts}\label{subsec:inference}
We will describe a technique for synthesizing procedure pre and postconditions.
To do that we will specialize the  static semantic rules and turn them into
inference rules. This will somehow resemble the process of developing type inference rules from a type system.


Figure~\ref{fig-calc-post} shows some of the rules to compute a postcondition $\calcPost{s}{p}$ from given $p$ and $s$. The complete calculus is given in~\cite{pest}.

\begin{figure}[ht!]
\begin{small}
$$
\begin{array}{c}
 \regla{\fuerza{p}{\safeExp{\sigma}(e)}}{\calcPost{\sentAssign{v}{e}}{p} = \expQuantif{\expExists}{v'}{\sust{p}{v}{v'} \land v = \sust{e}{v}{v'}}}{Q-ASSIGN}
 \\
 \\
\regla{\fuerza{p}{\safeBoolExp(g)}}
{\begin{array}{c}\calcPost{\sentIfThenElse{g}{s}{t}}{p} =\\ \clausuraExistencial{\calcPost{s}{p \land g} }{\localVars(s)} \lor \clausuraExistencial{\calcPost{t}{p \land \expNot{g}}}{\localVars(t)}\end{array}}{Q-IF}
\\
\\
\regla{\begin{array}{c}
\fuerza{\expTrue}{\safeBoolExp(inv)} \quad \fuerza{inv}{\safeIntExp(var)} \\
\fuerza{inv}{\safeBoolExp(g)} \quad \fuerza{p}{inv} \quad \fuerza{inv \land g}{var > \expNumeral{0}} \\
\calcPost{\sentAssign{var_{0}}{var}\;\; s}{inv \land g} = q' \\
\fuerza{q'}{inv} \quad \fuerza{q'}{var < var_{0}}\end{array}}
{\calcPost{\sentWhile{g}{inv}{var}{s}}{p} = inv\ \land\ \expNot{g}}{Q-WHILE}
\end{array}
$$
\end{small}
\caption{\pest postcondition calculus (fragment)}
\label{fig-calc-post}
\end{figure}

Observe the way in which these rules specialize those in Figure~\ref{fig-abstract-semantics}. For example, in the specialized rule for assignments, $\calcPost{\sentAssign{v}{e}}{p}$ is simply the strongest $q$ satisfying the original assignment rule.
The rule for \pestFuente{if} eliminates (using an existential closure\footnote{The existential closure of a boolean expression $b$ with respect to a set of variables $\{v_1,\ldots,v_n\}$  is $\exists v_1,\ldots,v_n(b)$}) the local variables defined in each of the branches from their respective inferred postcondition.

It is worth noticing that the inferred postcondition may contain unbounded existential quantifications. The main drawback of this is that, in that case, it is not valid to literally include it in the code. Observe also that an assertion with unbounded quantification cannot be checked at runtime; nevertheless, the inferred postconditions are correct by construction and the compiler can omit its associated runtime checks.

A similar approach can be used to infer procedure preconditions (denoted $\calcPre{s}{q}$). In this case, we use weakest-preconditions to find a suitable predicate $p$, except for \pestFuente{while} sentences where we simply use the provided invariant. The rules for computing $\calcPre{s}{q}$ can be found in~\cite{pest}.

Interestingly, we can infer non-trivial pre and postconditions even if the programmer provides no procedure annotations (other than loop annotations). First, we infer a predicate that guarantees the normal execution of a procedure body $s$:
$$
P\ \eqdef\ \calcPre{s}{\expTrue}
$$
This gives us a proper precondition for the procedure. Next, we strengthen $P$ by giving a symbolic initial value to each parameter ($p_i = \expVarAtPre{p_i}$) and use it to infer a postcondition $Q$:
$$
Q\ \eqdef\ \calcPost{s}{P \land p_1 = \expVarAtPre{p_1} \land \ldots \land p_k = \expVarAtPre{p_k}}
$$

The reader should verify that if the pre and postconditions of the procedures in Figures~\ref{fig:max} and~\ref{fig:arrayMax} are omitted, this procedure will infer logically equivalent ones. 


%

\subsection{Strengthening annotations}\label{subsec:strength}
We will now show how we can take advantage of \pest's restrictions on variable aliasing to incorporate inexpensive enhancements of invariants and postconditions. In a nutshell, we can propagate known facts about variables in certain scopes if we know that their contents are not altered. For example, let us say we know that before entering a loop it is always the case that variable $j$ has value $e$; then if we can determine that  the loop body does not update $j$, we can add $j = e$ to the invariant. 

In the absence of aliasing, a simple, sound approximation of the set of unmodified variables is to compute first the set of variables potentially modified and then take its complement. We call $\modVars(s)$ the set of (potentially) modified variables; its  definition is straightforward and can be found in~\cite{pest}.

What follows is a definition of our annotation strengthening function which takes a sentence $s$ and an entry point $p$ and produces a sentence $s'$ with possibly strengthened annotations. The translation does not alter assignments, local variable definitions or procedure calls. 
In the case of sentence sequence, we define it as:
$$
\trSent{s_1\ s_2}{p}\ \eqdef\ s'_1\ \trSentBig{s_2}{\calcPost{s'_1}{p}}
$$
where $s'_1 = \trSent{s_1}{p}$. That is, we first translate $s_1$ and use its postcondition to translate $s_2$.

When dealing with conditional statements we proceed by translating each branch using the conjunction of $p$ with the guard $g$:
$$
\trSent{\sentIfThenElse{g}{s_1}{s_2}}{p}\ \eqdef\ \begin{array}{l}\sentIfThenElse{g}{\trSent{s_1}{p \land g}\\}{\trSent{s_2}{p \land \expNot{g}}}\end{array}
$$

Finally, in the presence of a loop we strengthen its invariant. We existentially close $p$ over the variables that are potentially modified by the loop body:
$$
\trSent{\sentWhile{g}{inv}{var}{s}}{p}\ \eqdef\ \begin{array}{l}\sentWhile{g}{I}{var\\}{\trSent{s}{I \land g}}\end{array}
$$
where $I = inv \land \clausuraExistencial{p}{\modVars(s)}$.

\medskip
Using this technique we can strengthen both invariants and postconditions. For the latter case we simply take the postcondition of the last instruction of the procedure and existentially eliminate the local variables, in order to leave the postcondition only in terms of procedure parameters.

%

\section{High-level iteration constructs as annotations}\label{sec:constructs}
In this section we focus on what we call ``high-level iteration constructs''. These capture recurrent \emph{iteration patterns} and are frequently included in programming languages to reduce error-prone boilerplate code. Examples of such constructs are \textsc{Pascal}-style \emph{for}-loop and \textsc{C\#}'s \emph{foreach}. We will see next that this kind of constructs can be seen as implicitly carrying their own proof obligations, reducing the need for explicit annotations.

\begin{figure}[tb!]
	\centering
	\begin{tabbing}
		\=$\pmb{\mathbf{1 \leq i \leq \expArraySize{A}}}\ \land\ $\=$\expQuantifAcotada{\expForall}{0}{\le}{k}{<}{\pmb{\mathbf{i}}}{\phantom{cc}$\=$m \geq \expArrayAccess{A}{k}}\ \land\ $\=$\expQuantifAcotada{\expExists}{0}{\le}{k}{<}{\pmb{\mathbf{i}}}{\phantom{cc}$\=$m = \expArrayAccess{A}{k}}\ \land\ $\=$A = A@\mathrm{pre}$ \\
		\>\>$\expQuantifAcotada{\expForall}{0}{\le}{k}{<}{\pmb{\mathbf{\expArraySize{A}}}}{$\>$m \geq \expArrayAccess{A}{k}}\ \land\ $\>$\expQuantifAcotada{\expExists}{0}{\le}{k}{<}{\pmb{\mathbf{\expArraySize{A}}}}{$\>$m = \expArrayAccess{A}{k}}\ \land\ $\>$A = A@\mathrm{pre}$
	\end{tabbing}
	\vspace*{-1em}
	\caption{Invariant and postcondition of the main loop in the arrayMax procedure. Differences are highlighted.}
	\label{fig-inv-post}
\end{figure}

\subsection{The \emph{for} construct}
We first consider the \textsc{Pascal}-style \emph{for} loop and take as motivating example the \lstinline[emph={arrayMax}]{arrayMax} procedure in \S\ref{sec:motivation}. In Figure~\ref{fig-inv-post} we compare the postcondition of the loop with its invariant. Clearly, they are syntactically very close. In fact, if we are given a postcondition and a \emph{for}-loop, we can simply try to \emph{guess} a candidate invariant for the loop without even considering the loop body. Of course, the correctness of the invariant will have to be statically verified. On the other hand, a correct variant can be trivially obtained.

Formally, given a sentence of the form $\sentFor{i}{l}{h}{s}$ (without annotations) and its postcondition in the form of a predicate $Q_f$, we can macro-expand the sentence into the following \emph{lower-level} code:

\algoritmo{for}{14em}{
	\LocalDef{$i$}{\clasifInt}{$l$}
	\While{$i < h$}
		\Inv{$l \le i \le h\ \land\ \sust{Q_f}{h}{i}$}
		\Var{$h - i$}
	\Do
		\State $s$
		\Assign{$i$}{$i + \expNumeral{1}$}
	\Od
}

In order to apply this expansion we are required to provide a loop postcondition  $Q_f$. 
Nevertheless, this can be accomplished using the precondition calculus of \S\ref{subsec:inference} from the procedure postcondition up to the \emph{for}-sentence we need to expand. If $s$ contains a nested \emph{for}-construct, a postcondition for it can be derived from the inferred invariant and the expansion can be recursively applied.  

Observe that in the easyArrayMax procedure of Figure~\ref{fig:easyArrayMax}, $Q_f$ is simply the postcondition of the procedure, since the \emph{for} construct is located at the very end of the procedure body.


\subsection{Declarative iteration constructs}
\begin{figure}[bt!]
\begin{lstlisting}[emph=arrayInc]
arrayInc(A)
{
    local i <- 0
    while i < |A|
        :?! 0 <= i <= |A|
        :?! forall-k / 0 <= k < i : A[k] = A@pre[k] + 1
        :?! forall-k / i <= k < |A| : A[k] = A@pre[k]
        :# |A| - i
     do
         A <- update A on i with A[i] + 1
         i <- i + 1
     od
}
\end{lstlisting}
\caption{\pest procedure that increases each element in an array.}
\label{fig-array-inc}
\end{figure}

In \S\ref{subsec:inference} we saw that if the loop-invariants are provided, then postconditions can be omitted; conversely, in the previous section we showed that in the case of \emph{for}-loops invariants can be traded for postconditions. In this section we will show that when more specific iteration constructs are used, both can be dispensed with.  We will only consider here the \emph{map}-construct (reminiscent of the \emph{map} function over lists in functional languages) but the idea is easily extensible to other constructs.

Consider the procedure of Figure~\ref{fig-array-inc} (notice that the procedure contract was omitted and left for automatic inference). The loop iterates over an array $A$ using an indexing variable $i$. On each iteration, only the element of $A$ at position $i$ is updated. Furthermore, this is done taking into account only the value of $i$  and $A[i]$. The loop condition is not affected by changes to $A$. We call this iteration pattern, in which array elements are updated independently of the others, a \emph{map}. 

The invariant for this iteration pattern states that the  already visited array elements were updated whereas the remaining elements are unchanged. The variant reflects the fact that the array is traversed in a forward direction.

Using our proposed \pestFuente{map} construct, the procedure in Figure~\ref{fig-array-inc} can be written like this:

\begin{lstlisting}[emph={arrayInc}]
arrayInc(A) 
{
  map 
    A[i] <- A[i] + 1 
  in A[..i..]
}
\end{lstlisting}

\noindent Formally, this works as follows. Let $A$ be an array, $i$ a fresh variable and $s$ a sentence such that $\modVars(s) \smallsetminus \localVars(s) = \{A\}$ and $A$ is accessed only indexed by $i$. In that case, $\sentMap{s}{A}{i}$ is well-formed and gets expanded as follows:

\begin{lstlisting}
  local i <- 0
  while i < |A|
      :?! 0 <= i < |A|
      :?! forall-k / 0 <= k < i : $\sust{post_s}{i}{k}$
      :?! forall-k / i <= k < |A| : A[k] = A@pre[k]
      :# |A| - i
  do
      s
      i <- i + 1
  od
\end{lstlisting}
where $\mathit{post}_s = \calcPost{s}{0 \le i < \expArraySize{A}}$. It can easily be seen that the inferred invariant is always correct. Thus, not only can the compiler erase the assertion checks, it doesn't even need to statically verify them. 

Notice that the invariant inferred for the map high-level construct can be further strengthened using the technique presented in \S\ref{subsec:strength}.

\section{Experience}
In order to test these ideas we developed a tool called \budapest which is available online\footnote{http://lafhis.dc.uba.ar/budapest} as an Eclipse plug-in. 

The tool takes \pest programs, statically verifies them and compiles them to \textsc{Java} code.
For the verification step the code is translated into an intermediate assume/assert style language similar to BoogiePL ~\cite{boogie}. Unlike the Boogie verifier, our tool leverages the structure of the original program to discharge verification conditions one by one to SMT-solvers, such as CVC3~\cite{cvc-lite}, Yices~\cite{yices} and Z3~\cite{zap}. 
We split verification conditions in several pieces in order to allow the provers to work in parallel, so we can leverage each of the provers' power and combine their results. Splitting verification conditions also has the advantage of enabling early detection (during the verification process) of unsatisfiable conditions since once a subformula fails there is no need to continue with the rest of the subformulas.


We tested the usability of this tool in a first-year Computer Science course. The curriculum of the course includes proving correctness and termination of simple imperative programs specified using contracts written in a fragment of first-order logic and proven by way of Hoare axiomatic semantics~\cite{hoare}. With the aid of the tool they were able to automatically verify the correctness of their implementations of algorithms such as bubble and insertion sorting, linear and binary search, etc., which they previously had to do by hand.

The experience was very encouraging for the students since they found that concepts they had learnt during the course (preconditions, postconditions, invariants, variant functions) could be applied in ``real'' applications and produced code with guarantees of (partial) correctness and termination.

Finally, it is worth mentioning that the integrated nature of the \pest language forced the students to concurrently think about both the statements needed to implement an algorithm and the assertions required to prove its correction.

\section{Related Work}
There is a plenty of research on the automatic verification of programs. Due to lack of space we will just mention some of those we consider closer to our work.

There are several languages and tools that incorporate Hoare-style specifications to automatically prove partial correctness in imperative languages. Some well known examples are \textsc{ESC/Java}~\cite{esc-java}, \textsc{JML}~\cite{jml}, \textsc{ESC/Modula-3}~\cite{esc-modula}, \textsc{Spec\#}~\cite{spec-sharp-overview} and \textsc{SPARK/Ada}~\cite{spark}. These systems enrich the language with user provided annotations which can be checked on runtime or statically analyzed by generating verification conditions that are discharged to a theorem prover. 

The approach followed by \textsc{SPARK/Ada} is closer to ours in the sense that they impose limitations to the language in order to make verification possible. We believe they go too far since they limit syntactically to programs where assertions are decidable, making them too restrictive.
On the contrary, the problem with the other tools is that they deal with complex languages that provide many features such as polymorphism, concurrency, aliasing or reflection, which are comfortable for program development but complicate verification.

Our work follows the philosophy that a compiler (or interpreter) should try to reject misbehaving programs. 
This is similar the approach followed by several languages that provide expressive type systems such as \textsc{Deputy}~\cite{necula-deputy}, which uses dependent types~\cite{dependent-types} to type-check low level imperative programs. For functional languages there are many proposals; just to name one, Cayenne~\cite{cayenne} extends Haskell with dependent types. Recently,~\cite{htt} presented a type theory for higher order functional programs which incorporates Hoare style specifications into types, making it possible to enforce correct use of side effects.

\section{Final thoughts}
In this paper we presented \pest, a simple imperative programming language, which was designed with verifiability in mind. 
We proposed a series of techniques that aim at mitigating the annotation burden required to verify programs. Firstly, we showed an inference mechanism for procedure pre and postconditions and then a simple method to strengthen loop invariants and postconditions. 
As a distinguishing contribution we extended \pest with high-level iteration constructs that allow programmers to dramatically reduce annotations and yet maintain the ability to prove correctness of relatively complex pieces of software. 
Our first experiences showed that an integrated approach such as ours was gracefully adopted by young students with no pre-concepts with respect to programming languages or tools.

We would like to explore, in the near future, the possibility of extending \pest incorporating features that would increase its expressiveness without sacrificing verifiability. We plan to allow the programmer to define and use her own data types and provide means to reason about representation invariants. 
Adding dynamic memory support is another priority but, in order to keep a verifiable language, we believe we must enforce an alias control mechanism such as~\cite{noble-aliasprot}. 
Finally, we are interested in providing the means for programmers to define their own language constructs by feeding the \budapest tool with syntax and macro expansion definitions, and relying on the base tools (i.e. pre and postcondition calculi) that we provided in this paper to ensure correctness.

\bibliographystyle{plain} 
\bibliography{biblio}

\begin{thebibliography}{10}

\bibitem{cayenne}
L.~Augustsson.
\newblock {Cayenne-a language with dependent types}.
\newblock {\em ACM SIGPLAN Notices}, 34(1):239--250, 1999.

\bibitem{zap}
T.~Ball, S.~K. Lahiri, and M.~Musuvathi.
\newblock Zap: Automated theorem proving for software analysis.
\newblock In {\em LPAR}, 2005.

\bibitem{spark}
J.G.P. Barnes and Praxis Critical~Systems Limited.
\newblock {\em {High Integrity Ada: The SPARK Approach}}.
\newblock Addison-Wesley, 1997.

\bibitem{spec-sharp-overview}
M.~Barnett, K.R.M. Leino, and W.~Schulte.
\newblock {The Spec\# Programming System: An Overview}.
\newblock In {\em CASSIS}. Springer, 2005.

\bibitem{cvc-lite}
C.~Barrett and S.~Berezin.
\newblock {CVC L}ite: A new implementation of the cooperating validity checker.
\newblock In {\em Proceedings of the $16^{th}$ CAV}, 2004.

\bibitem{havoc}
S.~Chatterjee, S.K. Lahiri, S.~Qadeer, and Z.~Rakamaric.
\newblock {A reachability predicate for analyzing low-level software}.
\newblock 2007.

\bibitem{necula-deputy}
J.~Condit, M.~Harren, Z.~Anderson, D.~Gay, and G.C. Necula.
\newblock {Dependent Types for Low-Level Programming}.
\newblock {\em Lecture Notes in Computer Science}, 4421:520, 2007.

\bibitem{pest}
G.~de~Caso, D.~Garbervetsky, and D.~Gorín.
\newblock \textsc{Pest} formal specification.
\newblock Technical report, Universidad de Buenos Aires,
  \verb|http://lafhis.dc.uba.ar/budapest/| Theory section, 2008.

\bibitem{boogie}
R.~DeLine and K.R.M. Leino.
\newblock {BoogiePL: A typed procedural language for checking object-oriented
  programs}.
\newblock Technical report, Technical Report MSR-TR-2005-70, Microsoft
  Research, 2005.

\bibitem{esc-modula}
D.~L. Detlefs, K.R.M. Leino, G.~Nelson, and J.B. Saxe.
\newblock Extended static checking.
\newblock Technical Report \#159, Palo Alto, USA, 1998.

\bibitem{discipline-dijkstra}
E.W. Dijkstra.
\newblock {\em {A Discipline of Programming}}.
\newblock Prentice Hall PTR Upper Saddle River, NJ, USA, 1997.

\bibitem{yices}
B.~Dutertre and L.~de~Moura.
\newblock {The Yices SMT solver}.
\newblock {\em Available at http://yices.csl.sri.com/, August}, 2006.

\bibitem{daikon}
M.D. Ernst, J.H. Perkins, P.J. Guo, S.~McCamant, C.~Pacheco, M.S. Tschantz, and
  C.~Xiao.
\newblock The {Daikon} system for dynamic detection of likely invariants.
\newblock {\em Science of Computer Programming}, 2007.

\bibitem{esc-java}
C.~Flanagan, K.R.M. Leino, M.~Lillibridge, G.~Nelson, J.B. Saxe, and R.~Stata.
\newblock Extended static checking for {J}ava.
\newblock In {\em PLDI '02}, 2002.

\bibitem{vcs}
D.I. Good, R.L. London, and WW~Bledsoe.
\newblock {An interactive program verification system}.
\newblock In {\em Proceedings of the international conference on Reliable
  software table of contents}, pages 482--492. ACM New York, NY, USA, 1975.

\bibitem{hoare}
C.~A.~R. Hoare.
\newblock An axiomatic basis for computer programming.
\newblock {\em Commun. ACM}, 12(10):576--580, 1969.

\bibitem{jml}
G.T. Leavens, A.L. Baker, and C.~Ruby.
\newblock {JML}: {A} notation for detailed design.
\newblock In {\em Behavioral Specifications of Businesses and Systems}, pages
  175--188. Kluwer Academic Publishers, 1999.

\bibitem{leinoL05}
K.~Rustan~M. Leino and Francesco Logozzo.
\newblock Loop invariants on demand.
\newblock In {\em APLAS}, pages 119--134, 2005.

\bibitem{dependent-types}
J.~McKinna.
\newblock {Why dependent types matter}.
\newblock {\em Proc. ACM Symp. on Principles of Programming Languages (POPL
  2006)}, 2006.

\bibitem{de-millo}
R.A.~De Millo, R.J. Lipton, and A.J. Perlis.
\newblock Social processes and proofs of theorems and programs.
\newblock {\em Commun. ACM}, 22(5):271--280, 1979.

\bibitem{htt}
A.~Nanevski, G.~Morrisett, and L.~Birkedal.
\newblock Polymorphism and separation in hoare type theory.
\newblock In {\em ICFP '06}, 2006.

\bibitem{noble-aliasprot}
J.~Noble, J.~Vitek, and J.~Potter.
\newblock Flexible alias protection.
\newblock In {\em ECOOP '98: Proceedings of the 12th European Conference on
  Object-Oriented Programming}, pages 158--185, London, UK, 1998.
  Springer-Verlag.

\bibitem{pychecker}
N.~Norwitz.
\newblock {PyChecker}.
\newblock {\em SourceForge project http://pychecker. sourceforge. net}.

\bibitem{esc-haskell}
D.N. Xu.
\newblock Extended static checking for {Haskell}.
\newblock In {\em Proceedings of the 2006 ACM SIGPLAN workshop on {Haskell}},
  pages 48--59. ACM New York, NY, USA, 2006.

\end{thebibliography}

\end{document}